\documentclass{article}[12pt]
\usepackage{amsmath,amsfonts,amssymb,latexsym,ctable}
\usepackage{bbm}
\usepackage{caption}
\usepackage{subfig,graphicx}
\usepackage[utf8]{inputenc}
\usepackage{natbib}
\usepackage{bigints}
\usepackage{nameref}
\usepackage{hyperref}
\usepackage{amsthm}

\usepackage{geometry}
\newcommand{\Reals}{\mathbb{R}}

\newcommand{\E}{\mathbb{E}}

\DeclareMathOperator*{\pa}{pa}

\newcommand{\pn}[1]{\textcolor{blue}{{#1}}}
\newcommand{\lm}[1]{\textcolor{purple}{{#1}}}
\newcommand{\xib}{\mbox{\boldmath \(\xi\)}}
\newcommand{\sigmab}{\mbox{\boldmath \(\sigma\)}}

\newtheorem{proposition}{Proposition}
\usepackage{amsthm}

\newtheorem{definition}{Definition}[section]
\newtheorem{example}{Example}[section]

\title{Causal Discovery in Multivariate Extremes 
with a Hydrological Analysis of Swiss River Discharges}
   \author{L. Mhalla\footnote{Institute of Mathematics, EPFL, Switzerland. \textbf{Corresponding author}: linda.mhalla@epfl.ch},
    V. Chavez--Demoulin\footnote{HEC Lausanne, University of Lausanne and Expertise Center for Climate Extremes (ECCE), Switzerland}, and P. Naveau\footnote{Laboratoire des Sciences du Climat et de l’Environnement -- CNRS, France}}
\date{\today}

\begin{document}

\maketitle
\begin{abstract}
    Causal asymmetry is based on the principle that an event is a cause only if its absence would not have been a cause. From there, uncovering causal effects becomes a matter of comparing a well-defined score in both directions. Motivated by studying causal effects at extreme levels of a multivariate random vector, we propose to construct a model-agnostic causal score relying solely on the assumption of the existence of a max-domain of attraction. Based on a representation of a Generalized Pareto random vector, we construct the causal score as the Wasserstein distance between the margins and a well-specified random variable. The proposed methodology is illustrated on a hydrologically simulated dataset of different characteristics of catchments in Switzerland: discharge, precipitation, and snowmelt.
\end{abstract}
\section{Introduction}
\label{sec:intro}
\begin{figure}[!ht]
    \centering
        \includegraphics[scale=0.6]{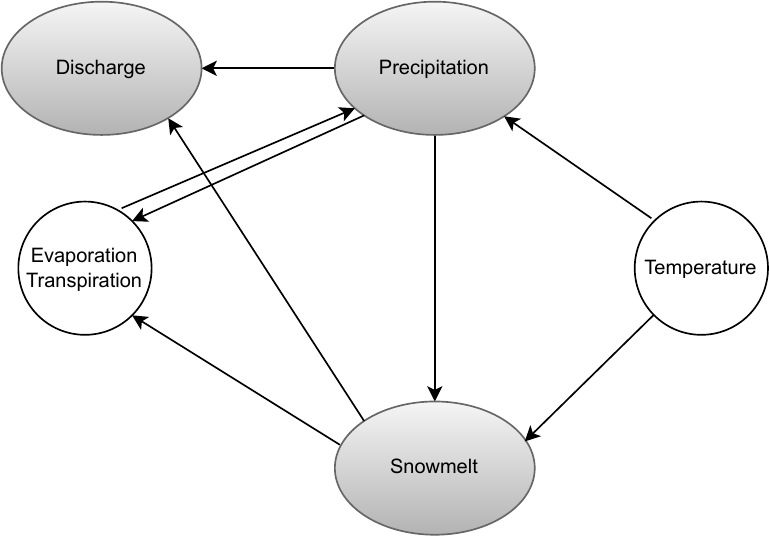}
    \caption{Potential atmospheric links  among  precipitation, snowmelt intensities and river discharges  in Switzerland  \cite[see, e.g.][]{Brunner2019,Schirmer23,Froidevaux2015}. Grey nodes represent the variables studied in this paper.}
    \label{fig:SwissGraph}
\end{figure}

The primary emphasis of causality has been on causal effects pertaining to 
averaged outcomes. 
Moving from averages towards extremal quantiles can be a necessary swift for various risk analysis. For example, studying possible causes of exceptional floods, record heatwaves and their associated high societal costs can be paramount for hydrologists, climatologists,  and re-insurance companies.  
Retrieving causal information from observational data, but at extreme levels poses a fundamental challenge across various scientific disciplines  \cite[see, e.g.,][for a  recent review]{chavezMhalla24}. 
For instance, in climatology,  the discipline of extreme events attribution investigates causal connections from  climate forcings such as greenhouse gases increases to observed responses on extreme phenomena like heatwaves or heavy rainfall
\cite[see, e.g.,][for a  review]{naveau_statistical_2020}. 
In this area,  causal generating mechanisms often exhibit distinct behaviours in the distribution bulk compared to its upper and lower tails.
For example, 
moderate rainfall are  influenced by factors such as prevailing wind patterns and orography. 
In the tails of the distribution, other atmospheric phenomena like atmospheric rivers 
\citep{Dettinger2011} can be added to the potential drivers of  extreme heavy precipitation. 
Hence, 
there is a clear need, at  least within  the climate and hydrological communities,  to develop simple and efficient causal tools for extremes. 
Besides a few recent theoretical advances \citep{Engelke2020, Gissibl2018}, very few studies deal,  in  an accessible manner to practitioners,  with the multivariate aspect of extremal causality. The goal of this study is to fulfill this gap. 
To motivate our particular setup,  
 the hydrological cycle of Switzerland will be the pedagogical thread. 
 The reason for such a choice is that this country has been well studied in terms of its hydrological  cycle \citep{Fatichi2015},  while  the complex orography of Switzerland represents a challenging test bed to discover how causal extreme links may vary spatially. 
    \label{fig:cycle}
In particular, 
snow plays a crucial role in Switzerland's water cycle. Over 30\% of its total precipitation falls as snow, and around 40\% of Swiss river runoff comes from snowmelt. Seasonal snow acts as a temporary reservoir for precipitation, releasing it in a condensed manner over a relatively short time, leading to variable seasonal runoff \citep{Schirmer23}.
These complex atmospheric potential  dynamical links  are schematically  summarized in Figure \ref{fig:SwissGraph}. 
From this connected graph, the main question is to identify the strengths of potential causal links that explain extreme river discharges. To address this issue, we first  need to recall  two statistical models that have  been particularly highlighted  in the recent causal literature dealing with extremes:  
linear structural causal models (LSCM) studied  by \cite{Gnecco_2021}
and recursive max-linear model (RMLM) introduced by \cite{Gissibl2018}. 
Under the LSCM setting, a causal framework among multiple variables, say  $\mathbf{Y}=(Y_1,\dots,Y_d)^\top$, is delineated by a directed acyclic graph (DAG), denoted $\mathcal{G}$, where the node set $V:=\{1,\dots,d\}$ denotes random variable indices    and the set of directed edges $E$ represents direct causal effects. 
The following relation
\begin{equation}
\label{eq:LSCM}
Y_j:=\sum_{k\in \pa(j)}\beta_{jk}Y_k+\varepsilon_j, \quad j\in V,
\end{equation}
  where $\pa(j)\subseteq V$ is the set of parents of $j$, $\beta_{jk}\in\Reals\setminus\{0\}$ is the causal weight of node $k$ on node $j$, and $\varepsilon_1,\dots,\varepsilon_d$ are jointly independent noise variables, defines a linear structural causal model (LSCM) with associated directed acyclic graph $\mathcal{G}$, in which the directed edge $(i,j)\in V\times V$ belongs to $E$ if and only if $i\in\pa(j)$. An extreme node observation $Y_j$ is the result of an extreme noise $\varepsilon_j$, or of a sum of weighted observations from the parents of $j$ in $\mathcal{G}$. 
  In contrast to the additive structure of a LSCM, a RMLM is based on the max-operator, i.e.,\
maxima over weighted parent nodes are considered. 
The main idea  is to propagate extremes throughout the max-linear equation defined as
\begin{equation} \label{eq:RMLM} 
Y_j := \max_{k \in \pa(j)} \max( c_{kj} Y_k, c_{jj} \varepsilon_j), \quad j\in V, 
\end{equation} with strictly positive weights $c_{kj}$ for all $j\in V$ and $k\in \pa(j)\cup {j}$.
The independent non-negative random variables  $\boldsymbol{\varepsilon}= (\varepsilon_1,\ldots,\varepsilon_d)^\top \in \mathbb{R}^d_{+}$ represent the vector of innovations. 
An extreme node observation $Y_j$ results from either an extreme  innovation $\varepsilon_j$ or from a large  maximum of some weighted observations from $Y_j$'s parents in $\mathcal{G}$.
A particular feature  of the model \eqref{eq:RMLM} is that the resulting joint distribution is discrete which may be  inconvenient for some  applications. 
Both models \eqref{eq:LSCM} and \eqref{eq:RMLM} can be expressed  with a DAG. 
In addition,  causal inference methods based on the LSCM \eqref{eq:LSCM} and RMLM \eqref{eq:RMLM} rely on the assumption of exogeneity, where  predictor variables are assumed to be independent of the error term.
This assumption is challenged by  the  presence of cycles like the ones observed in Figure~\ref{fig:cycle} that brings the of  identifying causality directions. 
To handle possible presence of feedback loops within a causal study, we propose a novel definition of causality at extreme levels. 
The proposed definition moves away from the concept of SEM and relies on a model-agnostic causal metric that is solely  based  on the assumption of the presence of a maximum domain of attraction \citep{Ferreira2014}. 
More precisely, leveraging  a representation of a  multivariate generalized Pareto (MGP) random vector \cite[see, e.g.][]{rootzen18}, we formulate the causal metric as the Wasserstein distance between  marginal distributions and a well-defined random variable.

The  paper is structured as follows. 
In Section \ref{sec:multi}, we review the basics of  MGP models  and recall some parametric examples that will be used in the simulation study. 
In Section \ref{sec:causalEx}, we propose a new definition of extremal causality in the bivariate setting. We assess our new causal methodology in Section \ref{seq:sim} and apply our method to the Swiss hydrological system in Section \ref{seq:app}. The paper ends by a conclusion   in Section \ref{sec:conclu}.
In terms of notations, the multivariate sample $(\mathbf{Y}_1, \ldots, \mathbf{Y}_n)^\top$  of size $n$ corresponds to independent and identically distributed (IID) real-valued copies of 
the $d$-dimensional random vector $\mathbf{Y}=(Y_1,\ldots,Y_d)^\top$ with associated realisations 
 realisation  $\mathbf{y}=(y_1,\ldots,y_d)^\top$.


\section{Multivariate Generalized Pareto model}
\label{sec:multi}
A fundamental aspect of univariate extreme-value analysis involves fitting a generalized Pareto (GP) survival distribution to a set of exceedances beyond a high threshold. This survival GP  distribution can be characterized by a positive scale parameter $\sigma$ and a shape parameter $\xi$ and it is equal to 
\begin{equation*}
    \overline{H}(z; \xi,\sigma)=\bigg(1+ \dfrac{\xi z}{\sigma}\bigg)^{-1/\xi}_{+},
\end{equation*} 
This two parameter expression allows for relatively straightforward statistical inference \citep{davisonSmith90}. In the  multivariate extremes context, the situation becomes complex as  there is not an unique way  to define a  multivariate extreme event. 
In addition, the family of distributions suggested by asymptotic theory no longer remains parametric
\citep{Ferreira2014}. 
Following \cite{rootzen06},  we define multivariate extremes  whenever, at least,  one component of  $\mathbf{Y}$ 
exceeds a large value. Such an extreme event  is denoted by $[\mathbf{Y}- \mathbf{u} \mid \mathbf{Y} \nleq \mathbf{u}]$ and has support on the $L$-shaped region $\{ \mathbf{x} \in \mathbb{R}^d : \vert \vert \mathbf{x} \vert \vert_{\infty} >0 \}$. 
We also assume that $\mathbf{Y}$ belongs to  the  domain of attraction of a max-stable distribution, i.e., 
if  $\mathbf{Y}_1, \ldots, \mathbf{Y}_n$  are independent copies of $\mathbf{Y}$, there exist sequences $\mathbf{a}_n \in (\mathbf{0}, \mathbb{\infty})^d$ and $\mathbf{b}_n \in \mathbb{R}^d$ such that
the distribution of the correctly re-normalized componentwise maxima defined by
\begin{equation*}
    \Pr \lbrace (\max_{i=1,\ldots,n} \mathbf{Y}_i - \mathbf{b}_n) / \mathbf{a}_n \leq \mathbf{y} \rbrace 
\end{equation*}
has a non-degenerate distribution as $n$ gets large. 
It has been shown by \cite{rootzen18} and \cite{Ferreira2014} that
the conditional vector $\mathbf{Y}- \mathbf{u} \mid \mathbf{Y} \nleq \mathbf{u}$ can then be  approximated, as $\mathbf{u}$ gets large,  by a random vector $\mathbf{Z}$ with a multivariate generalized Pareto (MGP) distribution 
 \cite[see][for a recent review on multivariate exteme value theory]{NaveauSegers24}. 
Any conditional margin of the  
multivariate GP distributed $\bf Z$ has a univariate GP in the sense that, for any $j=1, \ldots,d,$,
\begin{equation*}
    \Pr(Z_j > z \vert_j Z>0) = \overline{H}(z; \sigma_j,\xi_j), 
\end{equation*}
where $\sigma_j>0$ and $\xi_j\in \Reals$ corresponds to the marginal scale and shape parameters, respectively.
Any MGP vector $\bf Z$ with marginal parameters  $\xib=(\xi_1,\ldots,\xi_d)$ and  $\sigmab=(\sigma_1,\ldots,\sigma_d)$ can be rewritten as a   standardized version in the following way
\begin{equation}\label{Pareto}
\mathbf{Z} \overset{d}{=} \sigmab \dfrac{e^{\xib \mathbf{X}}-1}{\xib}, 
\end{equation}
with 
\begin{equation}\label{StandardPareto}
\mathbf{X}= E + \mathbf{U} - \max(\mathbf{U}), 
\end{equation}
where $E$ represents a univariate  exponential unit random variable and $\mathbf{U}$ any multivariate  random vector, independent of $E$ and $\max(\mathbf{U})= \max_{1\leq j \leq d}U_j$. The multivariate vector $\mathbf{X}$ is called a standard Pareto vector and, by construction, it support is 
 $\{\mathbf{x}\in \Reals^d: \mathbf{x} \nleq \mathbf{0}\}$ with  unit scale $\sigmab=\mathbf{1}$ and zero shape parameters $\xib=\mathbf{0}$.  
As there is no constraint on the choice of $\mathbf{U}$, the dependence structure in the vector $X$ is basically free and non-parametric by nature.   
In particular, if 
the random vector $\mathbf{U}$ has density $f_{\mathbf{U}}$ defined on $(-\infty,\infty)^d$, then the density of $\mathbf{X}$ can be expressed as 
$$h_{\mathbf{X}}(\mathbf{x};\mathbf{1};\mathbf{0})= \frac{\mathbbm{1}_{\{\max(\mathbf{x})>\mathbf{0}\}}}{e^{\max(\mathbf{x})}}\int_0^{\infty}f_{\mathbf{U}}(\mathbf{x}+ \log t)t^{-1}dt.$$
Another construction of standard MGP pdfs is due to \cite{rootzen18}. 
Suppose a $d$-dimensional random vector $\mathbf{T}$ with density $f_{\mathbf{T}}$ that satisfies $\E\left[e^{T_j}\right]< \infty $, for all $j=1,\ldots,d$, then a density of a GP distribution can be extracted as 
\begin{equation}\label{eq: f_T}
h_{\mathbf{T}}(\mathbf{z};\mathbf{1};\mathbf{0})=
\frac{\mathbbm{1}_{\{\max(\mathbf{z})>\mathbf{0}\}}}{\E \left[e^{\max(\mathbf{\mathbf{T}})}\right]}\int_0^{\infty}f_{\mathbf{T}}(\mathbf{z}+ \log t)dt.   
\end{equation}
This representation has the convenient property that any subvector $\mathbf{T}^{\prime}$ with at least one component above 0 of a GP random vector $\mathbf{Z}$ with density $h_{\mathbf{T}}$ is GP with same density $h_{\mathbf{T}^{\prime}}$ adjusted to the dimension of $\mathbf{T}^{\prime}$. 
\cite{Kiriliouk2018} provides  a review of constructions of GP vectors. 
From this work, we list below  three  parametric models that we will use in our simulation study.
\begin{example}[Logistic max-stable distribution]
\label{exple:logistic}
Let $\mathbf{W}\in \Reals^d$ be a random vector with independent Gumbel components with equal positive scale $\alpha$ and defined by 
$$
P(W_j\leq w) = \exp\{\exp(-\alpha w)\}.$$
If $f_{\mathbf{T}}$ is equal to $ f_{\mathbf{W}}$ but with  the restriction $\E[e^{T_j}] < \infty$, then 
\eqref{eq: f_T} leads
to the multivariare GP density with support  $\{\mathbf{z}\in \Reals^d : \mathbf{z} \nleq \mathbf{0}\}$ and associated logistic max-stable distribution. 
Note that in a similar way, the reverse Gumbel independent components lead to the multivariate GP distribution associated to the negative logistic max-stable distribution. 
\end{example}
\begin{example}[Dirichlet max-stable distribution]
\label{exple:dirichlet}
Same construction as in Example \ref{exple:logistic} but with $W_j$ following the pdf 
$$f_j(w) = \exp(\alpha_j w)\exp\{-\exp(w)\}/\Gamma(\alpha_j),$$
for $\alpha_j>0$ and $w \in (-\infty,\infty).$
\end{example}
\begin{example}[H\"usler–Reiss max-stable model]
\label{exple:HR}
Same construction as in Example \ref{exple:logistic} but with $\bf W$ following the pdf 
$$f_{\mathbf{W}}(\mathbf{w}) = (2\pi)^{-d/2}\mid \Sigma\mid^{-1/2} \exp\{-(\mathbf{w} -\mathbf{\beta})^T \Sigma^{-1}(\mathbf{w} - \mathbf{\beta})/2\}, $$
where $\mathbf{\beta} \in \Reals^d$ is the mean vector and $\Sigma\in \Reals^{d \times d}$ is the positive-definite covariance matrix. 
\end{example}
In the following section, we connect representation \eqref{StandardPareto} to our new definition of causality for extremes.



\section{MGPD causality}
\label{sec:causalEx}
We consider the bivariate setting of a random vector $(Y_1,Y_2)^\top$ which represents two nodes of a graph  like snowmelt and precipitation in Figure \ref{fig:cycle}. 
We suppose that the limiting tail behaviour of $(Y_1,Y_2)^\top$ is described by the MGP  vector  $\mathbf{Z}$  defined by \eqref{Pareto} and its related standard Pareto vector $\mathbf{X}$. 
To understand the dependence strength between  each component of $\mathbf{X}=(X_1,X_2)^\top$, we introduce the difference 
$V= X_1-X_2= U_1-U_2$ and we remark  that  Equation \eqref{StandardPareto} can be written   as
\begin{align}
  \begin{cases} 
    X_1 =  E + V-\max(0,V), \\
    X_2 =  E -\max(0,V), \label{eq:V}
  \end{cases}
\end{align}
where $V$ and $E$ are independent. 
From this system, we deduce that 
a absolute value of $V$ close to zero corresponds a strong dependence between $X_1$ and $X_2$, while a
large value of $|V|$ reflects almost  independence.  
In other terms,  the strength of dependence in $\mathbf{X}=(X_1,X_2)^\top$ is fully described by the random scalar $V$ and a strong (weak) departure  from zero reflects a weak (strong) dependence within $\bf X$. 
 Throughout the key role of  $\max(0,V)$ in  \eqref{eq:V}, 
 the (a)symmetry of $V$ will play a central role in our definition of extremal causality. 
 If the distribution of $V$ is strongly asymmetrical, say predominantly taking negative values, then $X_1=E+V$ and  $X_2=E$ will occur  more frequently than the alternative $X_1=E$ and  $X_2=E-V$. 
 On the other hand if the probability of $V$ being positive is higher than being  negative, than $X_1$ would be more  often equal to the unit exponential $E$. 
 Here, we argue that this asymmetrical feature  in the extremal dependence structure can be exploited to detect  extremal causality.
Intuitively, if the extremal causal structure in $(Y_1,Y_2)^\top$ is monotonic, then we would expect that an extreme event in $Y_1$ would always cause an extreme event in $Y_2$ (assuming $Y_1$ to be the parent of $Y_2$ in the $L$-shaped region), while the opposite does not necessarily hold. Thus, we expect $V$ to be more often negative than positive and $X_2$ to be closer to a unit exponential than $X_1$. Therefore, we propose to define the causally-induced asymmetry in the tails of $(Y_1,Y_2)^\top$ by comparing a  distance with respect to the unit-exponential $E$.  
Different distances exist to measure the proximity to a target pdf. 
In our the case, the Wasserstein distance leads to   direct computations with a simple interpretation in terms of the means of $X_1$ and $X_2$.
The Wasserstein distance \citep{Monge1781,Villani2008} is a metric between probability measures where one is interested in the ``minimal effort" of moving one (probability) measure to another. It is a particularly interesting proper metric when one of the probability measures is derived from the other with a small random perturbation. The Wasserstein distance between two univariate random variables $A$ and $B$    is defined as
\begin{equation}
    W_p(A,B) = \bigg\lbrace\int_0^1 \bigg(\vert F^{-1}(q) - G^{-1}(q) \vert \bigg)^p dq \bigg \rbrace^{1/p}, \notag
\end{equation}
with $F(x)=P(A\leq x)$ and  $G(x)=P(B\leq x)$. 
The special case $p=1$ being equivalent to
\begin{equation}
    W_1(A,B) = \int_\mathbb{R} \vert F(t) - G(t) \vert dt. 
    \label{eq:wasserp1}
\end{equation}
The following proposition (see proof in the \nameref{sec:appendix}) is the stepping stone for our proposed definition of extremal causality.
\begin{proposition}\label{prop.1}
Let $\mathbf{X}$ be a standard bivariate generalized Pareto vector  expressed as in \eqref{eq:V}. Then, the following equivalence holds
\begin{equation*}
    W_1(X_1, E) \geq W_1(X_2,E) \Leftrightarrow \E(X_1) \leq \E(X_2).
\end{equation*}
\end{proposition}
Therefore, a comparison of the Wasserstein distances between the scaled tail margins and the unit exponential distribution is informative about the asymmetry in the tail dependence through the sign of  $\E(X_i-X_j)$.
This is convenient as we have previously highlighted that the asymmetry in the difference $X_1-X_2$ was key in the structure of \eqref{eq:V}. Proposition \ref{prop.1} tells us that a simple difference in means is enough to capture the Wasserstein distance to the unit-exponential. 
In this context, we can now introduce a novel notion of extremal causality in the $d$-dimensional setting.
\begin{definition}
\label{def:causalEx}
Let $\mathbf{Y}=(Y_1,\ldots,Y_d)^\top$ be a $d$-dimensional vector with a joint distribution that is in the max-domain of attraction of a multivariate max-stable distribution. Denote by $\mathbf{X}$ its limiting standard generalized Pareto vector, and let 
\begin{equation}
s_{i \rightarrow j}=\frac{W_1(X_i, E) - W_1(X_j,E)}{\underset{k=1,\dots,d}{\max} \ W_1(X_k, E)}.
\label{eq:score}
\end{equation}
We refer to $s_{i \rightarrow j}$ as the causal score from the component $i$ to the component $j$.  
If $s_{i \rightarrow j}$ is finite and strictly positive, then 
we say that the component $i$ is the extremal cause of the component $j$. 
\end{definition}
Without this standardisation term, the score would be identical wherever both $X_1$ and $X_2$ are equally close to $E$ or far from $E$. 
In general, low scores indicate that   causality is  weak. 
In the case of non-identifiability,  e.g.\ where all the causes are excessively strong, then our score can be low.
The latter scenario aligns with a non-identifiable case according to our definition. This can be analogized to non-extreme causal discovery methods that utilize restricted additive noise models, wherein identifiability hinges on the presence of either a non-linear causal effect or non-Gaussian noise. 
This effect is illustrated on SEM relations in Figure~\ref{fig:simSEM}, when $\beta$ is large. We should note that although the definition relies on a bivariate score, extremal causal discovery in the multivariate setting is equivalent to finding the topological order of the graph. For instance,  the classical notion of conditional independence implies unconditional independence in max-stable vectors \citep{Ioannis2016}. Hence,  the causal graph associated to the standard Pareto vector $\mathbf{X}$ in~\eqref{StandardPareto} is fully connected. 
Relying on our proposed definition of extremal causality, the topological order is directly retrieved by ordering the Wasserstein distances $W_1(X_i,E)$.

\begin{figure}[!ht]
    \centering
    \includegraphics[scale=0.6]{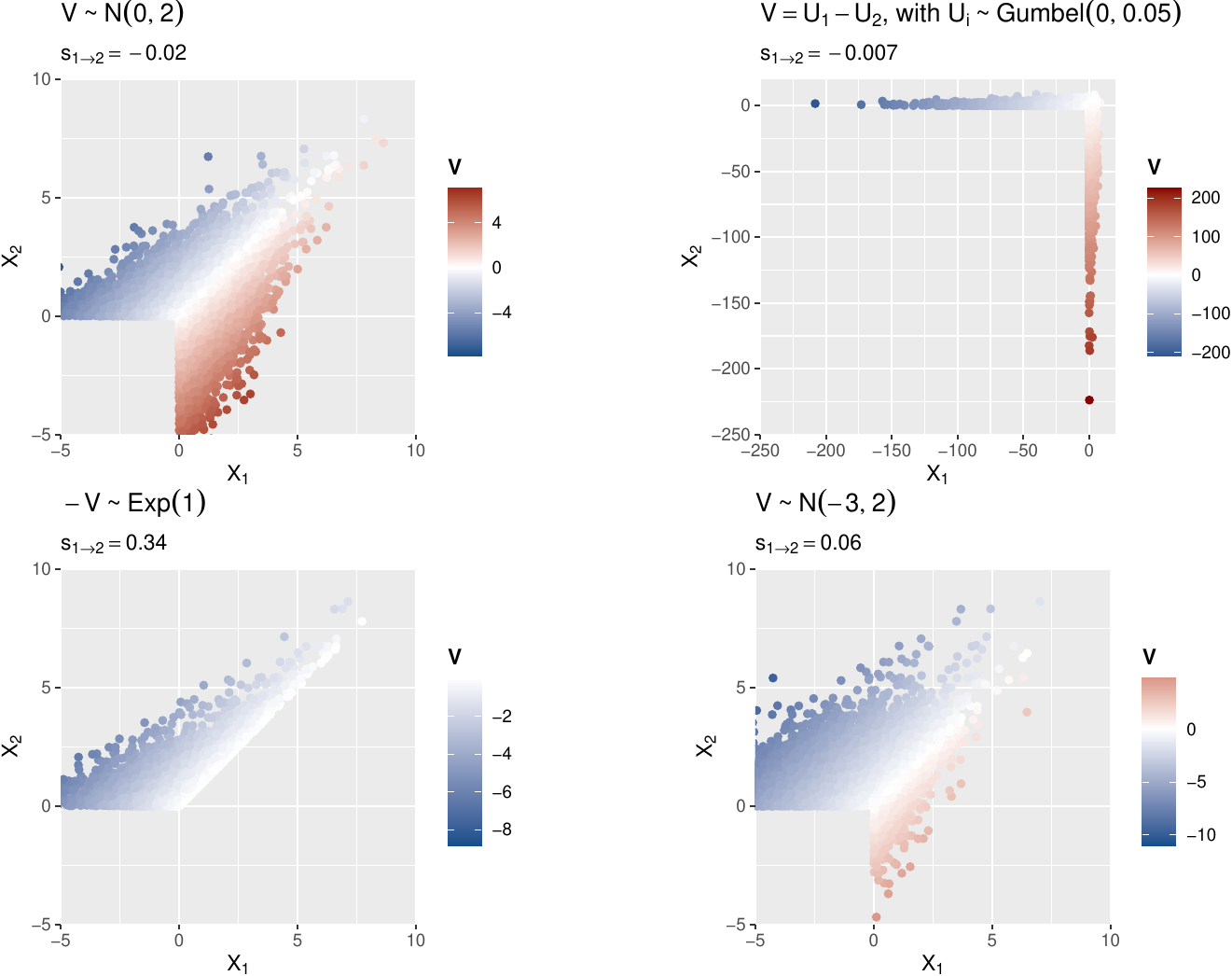}
    \caption{EGPD samples from \eqref{eq: f_T} with different distributions for the random variable $V$ reflecting different tail behaviours: symmetric with strong dependence (top left panel), symmetric with weak dependence (top right panel), asymmetric with strong dependence (bottom left), and asymmetric with moderate dependence (bottom right panel).}
    \label{fig:V}
\end{figure}

Figure \ref{fig:V} displays  $10^4$ bivariate  samples of $(X_1$  $X_2)^\top$ derived from various distributions of  $V=X_1-X_2$ highlighting the impact of its asymmetry around zero. 
The color of each point indicates the value of $V$.
For example, the two upper  panels showcase symmetrical  $V$, either following a zero-mean Gaussian (upper left) or  defiend as  the difference 
between two independent Gumbel distributed random variables (upper right).  
The  distributional symmetry is aligned with the absence of causal connection. 
In contrast, the two lower panels  correspond to asymetrical $V$, especially in the lower left corner.

\begin{figure}[!ht]
    \centering
    \includegraphics[scale=0.6]{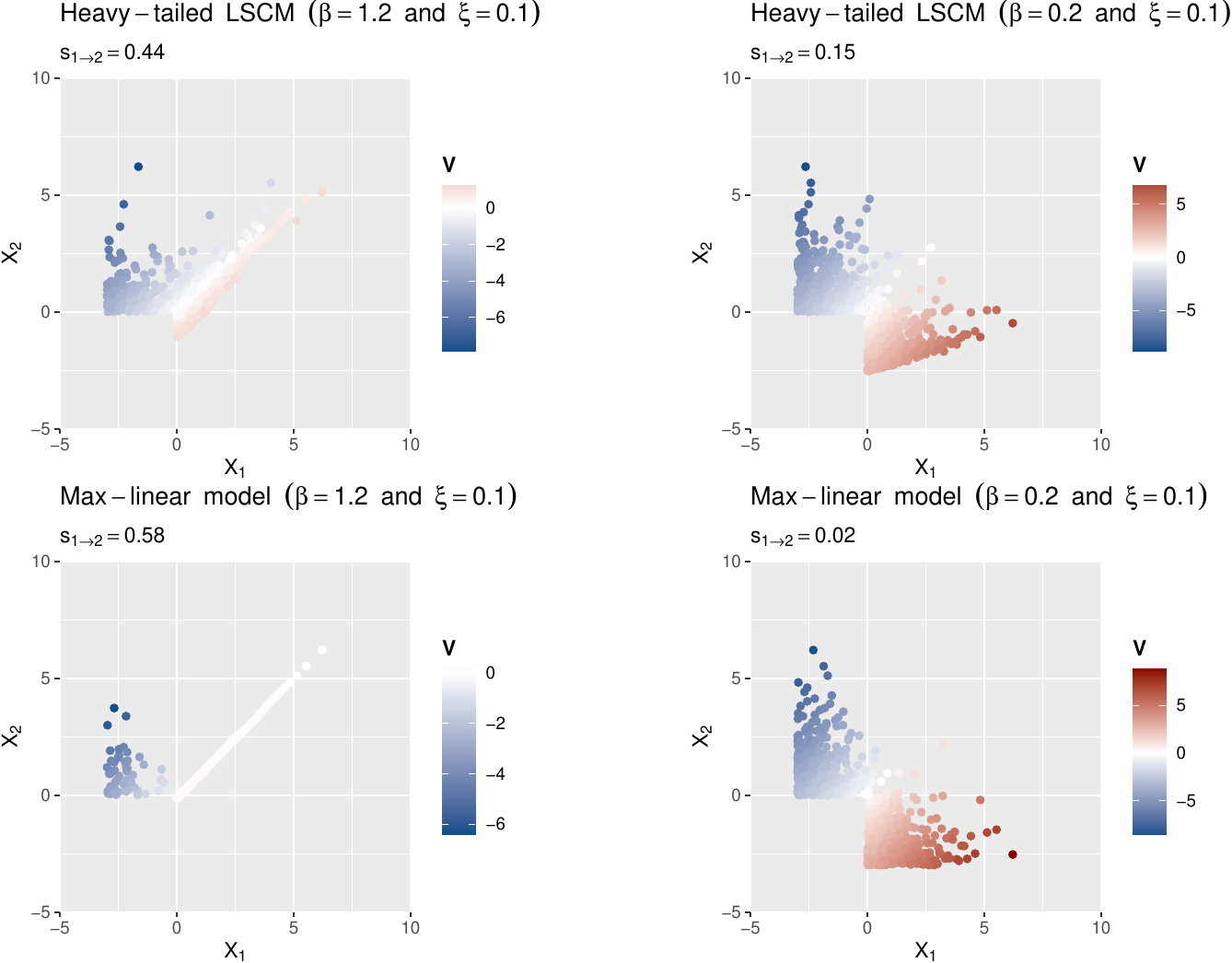}
    \caption{Bivariate samples from LSCM \eqref{eq:LSCM} and RMLM \eqref{eq:RMLM}. }
    \label{fig:SEM}
\end{figure}

Definition~\ref{def:causalEx} of extremal causality is broad. For instance, the SEM relations LSCM \eqref{eq:LSCM} and RMLM \eqref{eq:RMLM} with Pareto noise fall within this definition. That is, if two vertices are causally related in the heavy-tailed SEM, then the SEM-associated parent is the \textit{extremal cause} of the SEM-associated child. Following our definition of extremal cause, then $V$ is asymmetric with more negative values on the child axis. Figure~\ref{fig:SEM} illustrates this situation where $Y_1$ is the parent of $Y_2$ in the LSCM \eqref{eq:LSCM} (top panels) and RMLM \eqref{eq:RMLM} (bottom panels) with $\beta=1.2$ (left) and $\beta=0.2$ (right) and Pareto noise $\varepsilon_j$, $j=1,2$ with shape parameter $\xi=0.1$ in all these heavy-tailed cases. The asymmetry of $V$ is more pronounced for large values of $\beta$ as all extremes generated by $Y_1$ (or equivalently by $\varepsilon_1$) are also extremes of $Y_2$ but $Y_2$ also generates its extremes through $\varepsilon_2$ as $\beta$ is close to 1. As a consequence of this, the score \eqref{eq:score} is higher for larger values of $\beta$. 
Causal links encoded in a heavy-tailed LSCM or RMLM remain valid at extremal levels and according to our definition, higher (absolute) values of \eqref{eq:score} reflect stronger causal links, though the relation between the two is mediated by the tail index of the noise, as assessed by a simulation study in the next section. 

To summarize, our definition of extremal causality stems from the strength and
asymmetry of the extremal dependence structure. Specifically, when extremal dependence exhibits both strength and asymmetry, it indicates a potent extremal causal connection, even in the absence of structural causal relationships between variables. On the other hand, when extremal dependence is asymmetric but weak, it suggests a less influential extremal causal link. In cases where extremal dependence is either strong or weak but symmetrical, it signifies the absence of an extremal causal connection altogether.

\section{Simulations}
\label{seq:sim}
\begin{figure}[!ht]
    \centering
    \includegraphics[scale=0.7]{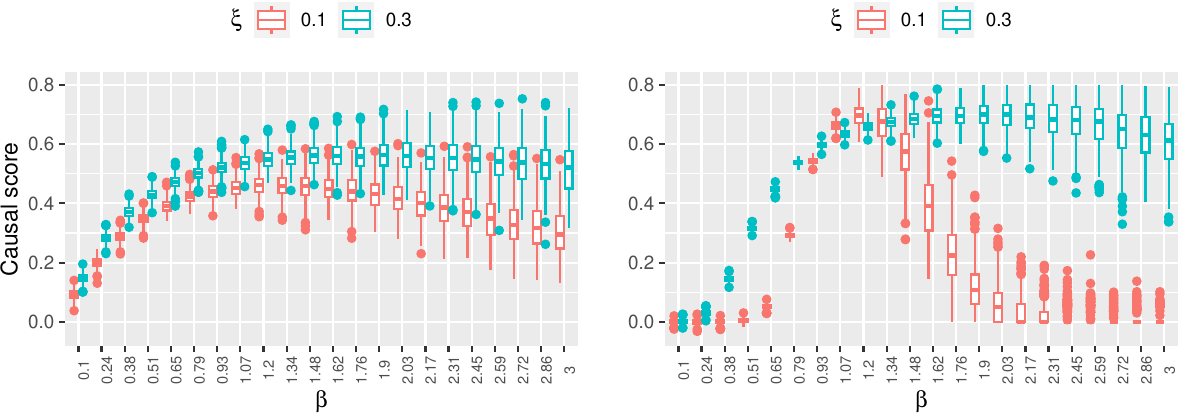}
    \caption{Score \eqref{eq:score} $s_{1 \rightarrow 2}$ based on simulated data from LSCM \eqref{eq:LSCM} (left) and RMLM \eqref{eq:RMLM} (right). Results rely on $500$ bootstrap replicates.}
    \label{fig:simSEM}
\end{figure}

We ran different simulation studies to assess our extreme causality method in various contexts. The first is based on the SEM relations LSCM \eqref{eq:LSCM} and RMLM \eqref{eq:RMLM}. The heavy-tailed LSCM and RMLM from which we simulated $n=10^4$ data and assumed a MGPD above a marginal threshold at the $95\%$ quantile are respectively

$$ \text{LSCM:} \left\{ \begin{array}{ccl}
 Y_1 & =  & \varepsilon_1, \\
  Y_2 & = &  \beta Y_1 + \varepsilon_2, 
  \end{array}
 \right.   $$
and
$$ \text{RMLM:} \left\{ \begin{array}{ccl}
 Y_1 & =  & \varepsilon_1, \\
  Y_2 & = &  \max(\beta Y_1, \varepsilon_2), 
  \end{array}
 \right.   $$
 both with $\varepsilon_j \sim$ Pareto with shape parameter $\xi = 0.1$ and $\xi = 0.3$  and with $\beta$ varying from 0.1 to 3. The boxplots of the resulting causal scores \eqref{eq:score}, represented in Figure~\ref{fig:simSEM}, lay all on the strictly positive side for all values of $\beta$. For low values of $\beta$, causality exhibits reduced strength in both SEM relations, consequently yielding smaller scores. At the uppermost values of $\beta$, for a small shape parameter $\xi$, the SEM relations lean towards perfect dependence, rendering causal discovery more challenging, especially for RMLM, but still achievable. 

\begin{figure}[!ht]
    \centering
    \includegraphics[scale=0.7]{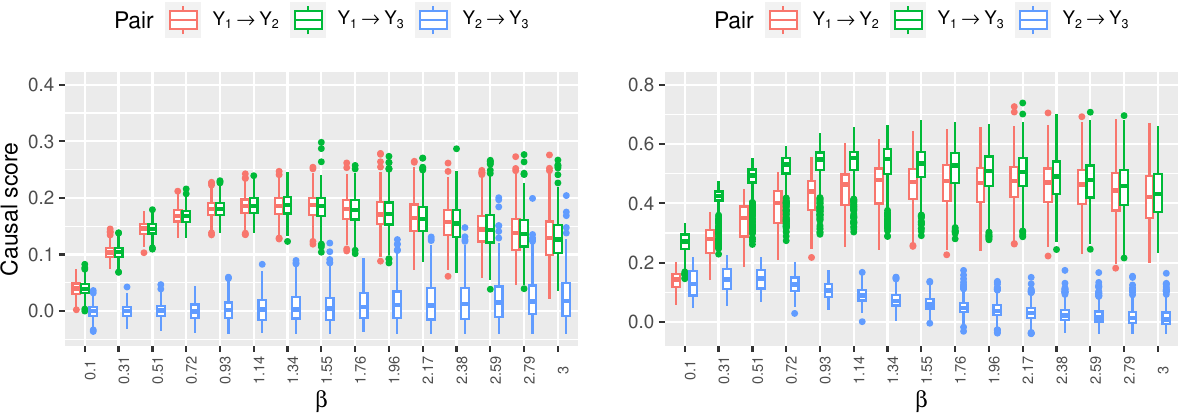}
    \caption{Score \eqref{eq:score} based on simulated data from LSCM \eqref{eq:LSCM} with confounder and no direct causal link (left) and with direct link (right). Results rely on $500$ bootstrap replicates and the shape parameter $\xi$ is set at $0.1$.}
    \label{fig:simSEM_confounder}
\end{figure}

Building on the LSCM structure, we now consider two settings with a confounder variable $Y_1$ and the absence or presence of a direct causal link between the variables of interest $Y_2$ and $Y_3$. Precisely, the former is 

$$ \left\{ \begin{array}{ccl}
 Y_1   &=& \varepsilon_1, \\
 Y_2 & =  & \beta Y_1 + \varepsilon_2, \\
 Y_3 & = &  \beta Y_1 + \varepsilon_3, 
  \end{array}
 \right.   $$
and the latter is 

$$ \left\{ \begin{array}{ccl}
 Y_1   &=& \varepsilon_1, \\
 Y_2 &=& \beta Y_1 + \varepsilon_2, \\
 Y_3 &=&  \beta Y_1 + \gamma Y_2 + \varepsilon_3, 
  \end{array}
 \right.   $$
where $\gamma$ is a random coefficient drawn between $0.1$ and $3$. The boxplots of the resulting causal scores \eqref{eq:score} are represented in Figure~\ref{fig:simSEM_confounder}. When $Y_2$ and $Y_3$ are not causally related, the presence of the confounder $Y_1$ has no impact on the causal score $s_{2 \rightarrow 3}$ which is centered around zero. The causal scores $s_{1 \rightarrow 2}$ and $s_{1 \rightarrow 3}$ behave similarly to the LSCM setting in Figure~\ref{fig:simSEM} (left panel), as expected. In the presence of a direct causal link between $Y_2$ and $Y_3$, $s_{1 \rightarrow 3}$ is on average higher than $s_{1 \rightarrow 2}$ due to the presence of two causal pathways between the variables $Y_1$ and $Y_3$. As the causal coefficient $\gamma$ is randomly chosen, the causal score $s_{2 \rightarrow 3}$ seems to be pivoted by the coefficient $\beta$ and is strictly positive for small values of $\beta$ and decreases rapidly to zero with increasing $\beta$.

 \begin{figure}[!ht]
    \centering
    \includegraphics[scale=0.7]{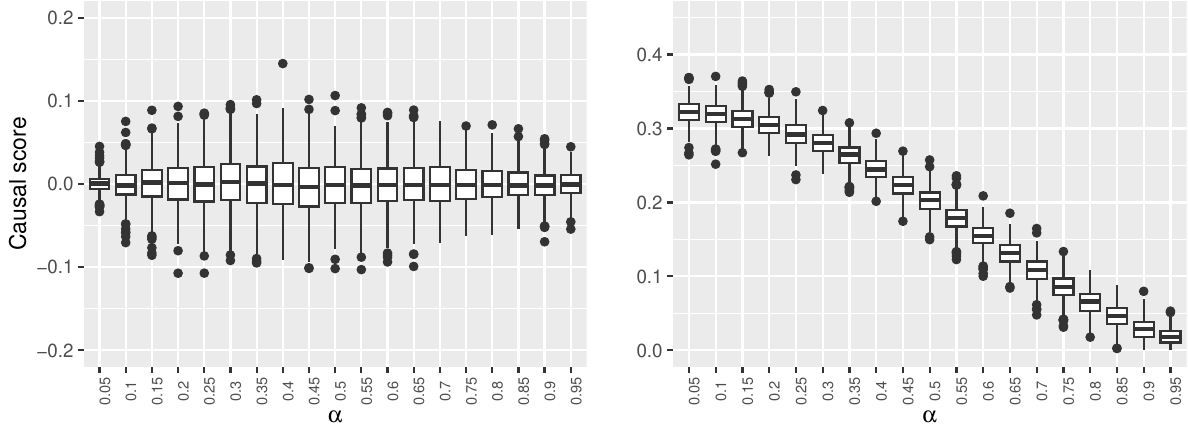}
    \caption{Score \eqref{eq:score} based on simulated data from logistic model with varying dependence parameter (left) and from asymmetric logistic model with varying dependence parameter (right). Results rely on $500$ bootstrap replicates.}
    \label{fig:asymLog}
\end{figure} 
We now run logistic model-based simulations. They are based on simulating $n=10^4$ data from a logistic or an asymmetric logistic extreme value copula. The bivariate asymmetric logistic copula introduced by \cite{Tawn1990} is 

$$C_{\alpha,\beta_1,\beta_2}(u,v)= \exp\left[
-\left\{
(-\beta_1 \log u)^{1/\alpha}+ (-\beta_2 \log v)^{1/\alpha}
\right\}^{\alpha}+(1-\beta_1)\log u + (1-\beta_2)\log v
\right]$$
with $\beta_1,\beta_2 \in \left[0,1\right]$, the asymmetry parameters. The case $\beta_1=\beta_2=1$ defines the (symmetric) logistic copula.  The parameter $\alpha \in (0,1]$ specifies the strength of dependence with values close to zero corresponding to strong dependence and values close to one, to independence. 
For both the logistic and asymmetric logistic cases, we apply our extremal causality method on the MGPD distributed data above their marginal 95\%-quantile. Figure \ref{fig:asymLog} shows the resulting scores for the logistic data (left panel) and for the asymmetric logistic data with stronger asymmetry parameter ($\beta_1=0.8$) for the first component than for the second ($\beta_2=0.2$)(right panel) against different values of the tail dependence parameter $\alpha$. 

As expected, the score is close to zero in case of the symmetric logistic model and positive for the asymmetric model. The positiveness adequately suggests that the first component is the cause of the second. Again, the higher is the tail dependence (or equivalently the smaller is the value $\alpha$), the stronger is the evidence for extremal causality. Our causal score can be related to the asymmetric tail Kendall's $\tau$ introduced by \cite{deidda} where it is shown how it can be used to inform the direction of causality between the extreme observations that present asymmetric tail dependence structure. The findings from their simulated asymmetric copula and from ours for this model are aligned. 

\section{Extremal analysis of Swiss hydrological catchments}
\label{seq:app}
During the Spring season, in particular, extreme precipitation events can directly lead to increased surface runoff and contribute to higher water levels in rivers and streams. These events can also result in saturation of soils, leading to increased infiltration and groundwater recharge, which can contribute to higher baseflow in rivers over time. In Switzerland, mountain regions have significant snowpack and extreme snowmelt can occur due to rapid warming or rain-on-snow events in Spring. Rapid snowmelt can in turn lead to increased surface runoff and contribute to higher river discharges, particularly in Spring when snowpacks are melting. Extreme discharges in rivers and streams are therefore often the result of a combination of factors, including extreme precipitation and rapid snowmelt. In the Plateau, atmospheric processes and topographical features can lead, under conditions such as condensation, to situations where precipitations are due to snowmelt. 

We consider data simulated using the hydrological modelling system PREVAH (PREecipitation-Runoff-EVApotranspiration Hydrotope model) (\cite{Brunner2019}; \cite{Viviroli2009}). The dataset consists of 307 catchments in Switzerland for which discharge, precipitation,
and snowmelt were simulated at a daily-resolution from 1981 to 2016. The catchments' flood events are mainly driven either by snowmelt (Alps)
or rainfall (Jura, Plateau, and Southern Alps) or by their mixture
(Pre-Alps), see \cite{Froidevaux2015}. The system of the hydrological variables is spatially dynamic and the goal is to assess the extremal causal mechanisms over all
catchments during Spring (March--April--May). The atmospheric dynamics between the studied variables are summarized in Figure \ref{fig:SwissGraph}.

\begin{figure}[!ht]
    \centering
    \includegraphics[scale=0.55]{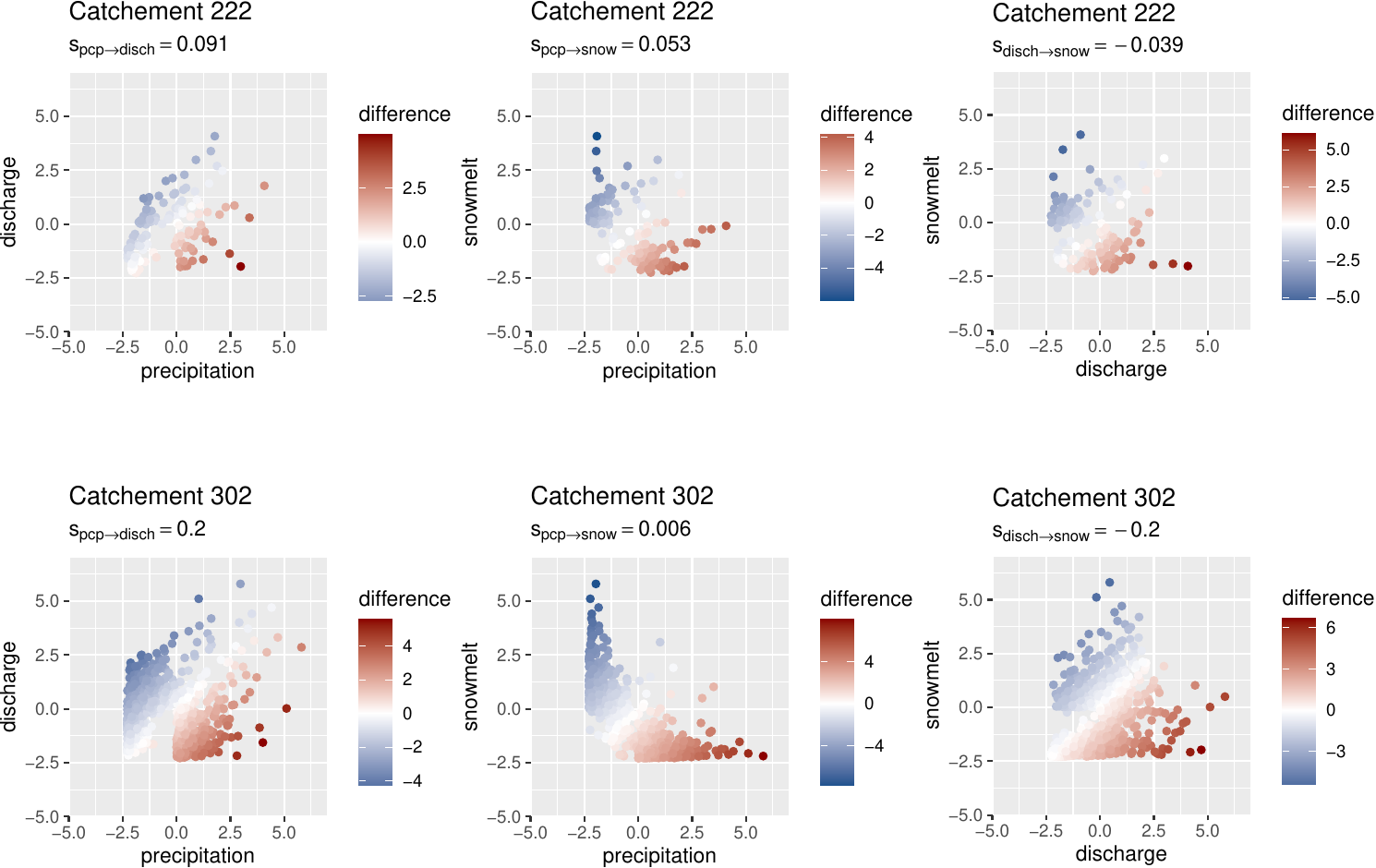}
    \caption{Pairs of variables transformed empirically to the standard Pareto scale, at two different catchments. Colouring scheme reflects the magnitude of the difference between the displayed margins.}
    \label{fig:plot_biv_application}
\end{figure}

Understanding the causal dynamics between the three factors precipitation, discharge, and snowmelt at their extreme level is crucial for effective water resources management, flood risk assessment, climate change adaptation efforts, and for accurately predicting and mitigating the impacts of extreme hydrological events. Switzerland is a relief country where overall, altitude influences the spatial distribution of precipitation, the duration and timing of snow accumulation and melt, and the characteristics of river basins, all of which play critical roles in shaping the causal dynamics between extreme precipitations, extreme snowmelt, and extreme discharges in mountainous regions. The objective of our study is the evaluation of extreme causal mechanisms between the three variables when at least one of the three is extreme, that is higher than its $90\%$ empirical marginal quantile, a situation that we will call ``under extreme condition". To remove the time lag effects and temporally align the variables without imposing a directional bias on their causal dynamics, we pre-processed the data such that marginal events that might result in flooding are contemporaneous. For instance, we consider moving windows of three days\footnote{Results presented in this analysis were not sensitive to this choice of the length.} over which accumulated precipitation, average snowmelt, and discharge level at the central point are computed. 
This results in a dataset of 3309 observations. While the pre-processing aims at aligning potential isolated or compound extreme events, we need to ensure that catchments experienced snowmelt over the considered time frame, i.e., simulated values for this variable can be non-zero. This was not, for instance, the case for three out of the 307 catchments, where all simulated snowmelt values during the March--April--May period were zero. For these catchments that have small areas, though not the smallest in the dataset, we exclude snowmelt from the causal discovery in extreme events. For the rest of the catchments where non-zero values of snowmelt are present, we remove instances where no snowmelt was observed and perform causal discovery for the triplet of variables. This way, we avoid biasing our results with observations under an extreme condition but with a complete absence of snowmelt. Figure~\ref{fig:plot_biv_application} displays the pairs of variables on the standard Pareto vector scale and at two different catchments.

\begin{figure}[!ht]
    \centering
    \includegraphics[scale=0.6]{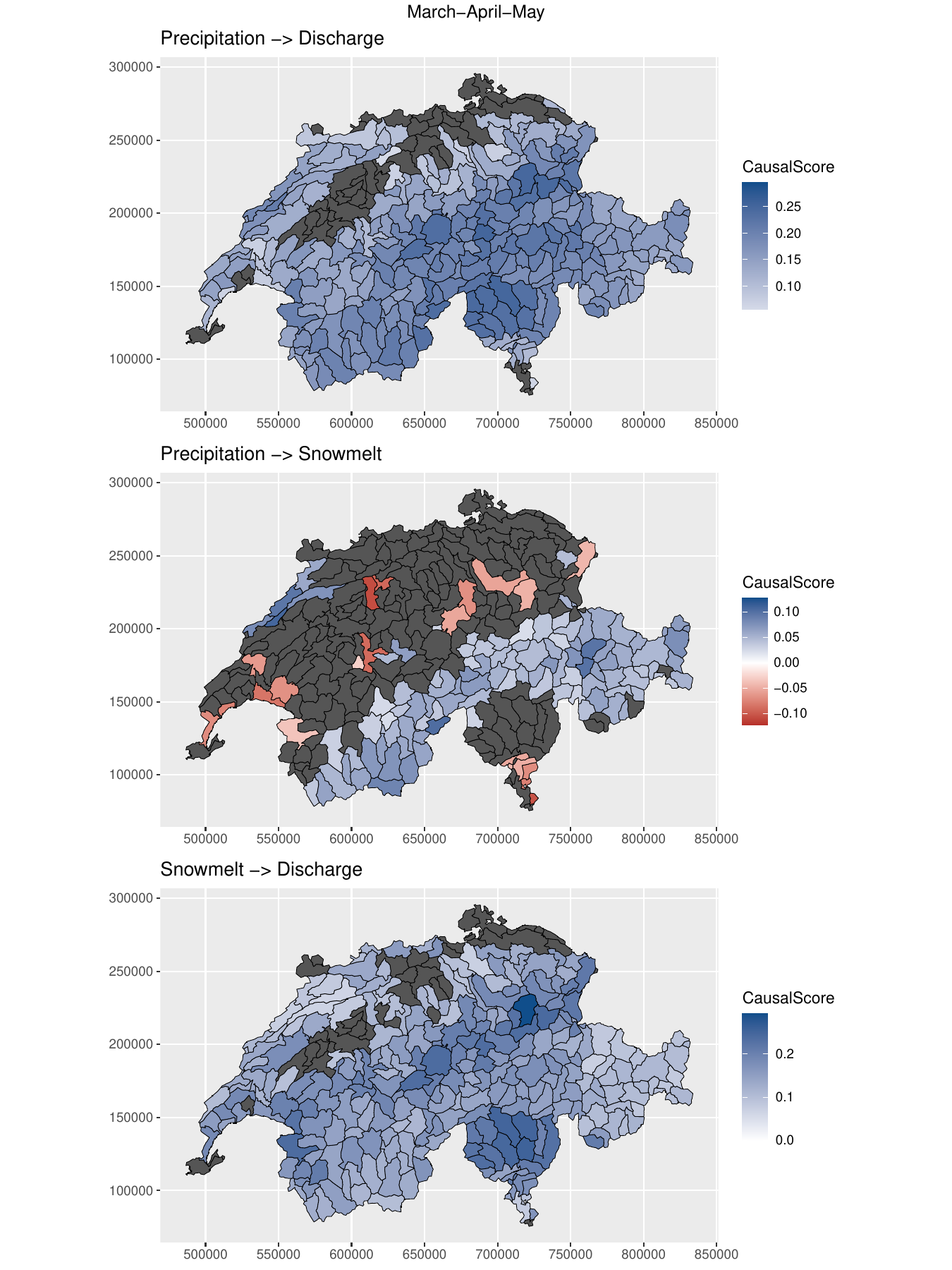}
    \caption{Pairwise causal score \eqref{eq:score} for the three studied variables: precipitation, snowmelt, and discharge. Positive scores (in blue) highlight significant causal link suggested, negative scores (in red) represent the inverse causal link suggested, and dark grey areas show no significant causal link.}
    \label{fig:SwissMaps}
\end{figure}

For each of the 307 catchments over Switzerland, we consider events under extreme condition and perform our causal discovery method. That is, for each pair of variables considered, we compute the causal score \eqref{eq:score} and retain significant score only when zero is not in the 95\%- confidence interval obtained from bootstrapping the data 300 times. The three panels of Figure~\ref{fig:SwissMaps} show the resulting values of the score over the different regions of Switzerland in Spring (March--April--May). The top panel highlights the (expected) causal effect of precipitation on discharge under extreme condition. The scores are higher at altitude. Indeed in mountainous regions, the interaction between extreme precipitation and extreme discharge in Spring can be enhanced by extreme snowmelt situations in areas where snowpack accumulation can reach significant depths. 
The middle panel shows significant causes of precipitation on snowmelt under extreme condition at altitude only. During Spring, warm fronts may bring rain to higher elevations where the snowpack exists. This rain falling onto the snowpack can accelerate the melting process. Rain-on-snow events are particularly significant because rainwater has a higher density than snow, so it can infiltrate the snowpack more easily and increase the rate of melt.
The map shows many areas in dark grey with no significant causal links between precipitation and snowmelt especially in the Plateau. That may be due to the lack of snow in these areas. 
The red areas highlight situations where snowmelt causes precipitation. This happens at low elevations in the Plateau in Switzerland (negative scores) and is primarily due to atmospheric processes and topographical features. When warm, moist air masses move across the region, they are forced to rise as they encounter mountain ranges. This process is known as orographic lifting. Another explanation is condensation. As the air rises, it cools due to decreasing atmospheric pressure at higher altitudes. Cooler air holds less moisture, leading to condensation of water vapor and the formation of clouds. Eventually, the condensed water droplets coalesce and fall as precipitation. 
Additionally, during Spring, as temperatures rise, snowmelt occurs in the higher elevations of the Alps and the Jura Mountains. This melted snow adds moisture to the atmosphere in the form of water vapor. As the air masses move over the higher elevations, they pick up moisture from the melting snow and become saturated with water vapor. When these moist air masses descend on the leeward side of the mountains, they encounter the lower elevations of the Swiss Plateau. As the saturated air descends from higher altitudes to lower elevations, it undergoes adiabatic compression, causing it to warm. However, the air retains its moisture content. When the warm, moist air reaches lower elevations, it cools again due to mixing with cooler surface air. This cooling can lead to further condensation and precipitation, even at lower elevations on the Plateau. The Swiss Plateau's relatively flat terrain allows moisture-laden air masses to spread out and distribute precipitation over a wide area, enhancing the likelihood of precipitation occurring at low elevations. Overall, the combination of orographic lifting, snowmelt-induced moisture, adiabatic processes, and the topographical characteristics of the Swiss Plateau contribute to the occurrence of precipitation at low elevations during periods of snowmelt in the higher mountain regions. This phenomenon is an example of how atmospheric and geographical factors interact to influence local weather patterns and precipitation distribution under extreme condition.
The bottom panel of Figure~\ref{fig:SwissMaps} shows a rather uniform effect of snowmelt on discharge under extreme condition across the regions during Spring likely resulting from combination of geographical features, climatic conditions, hydrological connectivity, and water management practices in Switzerland. Finally, we address the grey catchments where no significant causal link was found. Although grey catchments in the middle panel might be a consequence of the intricate atmospheric relations linking precipitation and snowmelt, those in the top and bottom panel seem to occur only in low altitude catchments and mostly in the Plateau. These are catchments where snowmelt volumes are relatively low and soil moisture is very high \citep{Brunner2019}. As soil moisture has been found to play an important role in runoff generation of floods \cite{Berghuijs}, a possible explanation for failing to uncover causal links might be the absence of a major atmospheric actor that is the soil moisture. This partial explanation should be supplemented by a thorough analysis of the raw simulated data and/or their calibration process, in the catchments where causality is unidentifiable in all pairs.

\begin{figure}[!ht]
    \centering
        \includegraphics[scale=0.8]{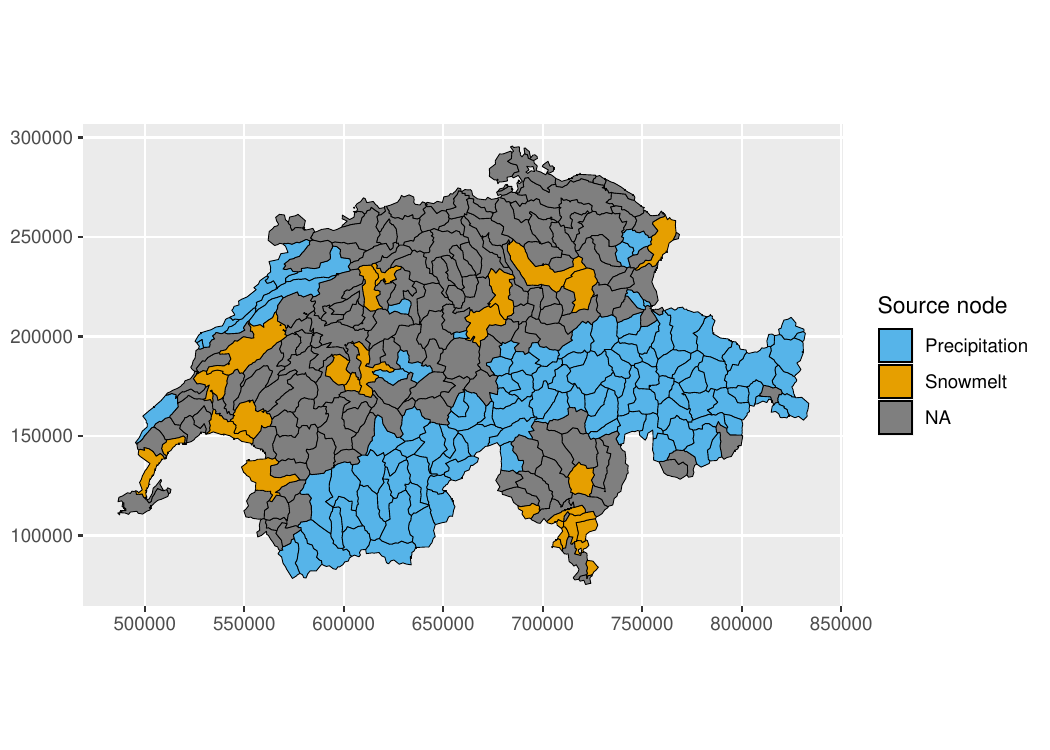}
    \caption{Source node based on ordering the Wasserstein distance \eqref{eq:wasserp1} among the three variables and based on majority vote among 300 resampling calculations. Dark grey highlights undefined majority vote.}
    \label{fig:Source}
\end{figure}

By ordering the Wasserstein distance \eqref{eq:wasserp1}  among the three variables, we can discover the source node over Switzerland under extreme condition. Contrarily to \cite{Gnecco_2021} where the EASE algorithm requires a sequential source node discovering, our method  can be  achieved in one step. This is because our transformed variables are all on the same scale and comparable as the term $\max(U_1,\ldots,U_d)$ of \eqref{StandardPareto} appears in the expression of all transformed margins. Figure~\ref{fig:Source} shows the dominating source nodes over the 307 catchments. To assess uncertainty of our methodology, data are bootstrapped 300 times and we retain a variable as a source node only when this is the case at least 95\% of the time. In accordance with findings in Figure~\ref{fig:SwissMaps}, precipitation dominates on relief and snowmelt in some low-altitude catchments. These catchments, located in the Plateau, the Pre-alps, and Southern Alps, are all adjacent to major lakes in Switzerland. The fact that snowmelt becomes the primary driver of discharge and precipitation dynamics, when extreme conditions occur, might be related to this proximity to water sheds. This can also be explained by the fact that the country's mountainous terrain and high elevation regions accumulate substantial snowpack during winter, which rapidly melts as temperatures rise. This accelerated snowmelt contributes significantly to river flow, aided by Switzerland's steep slopes and narrow valleys, facilitating rapid water transfer.

\section{Conclusion}
\label{sec:conclu}
The Earth system is composed of many intricate linkages that make causal discovery a challenging task. This work addresses discovery of causal links at extreme levels between three hydrological variables in different catchments in Switzerland. Relying on the multivariate extreme value theory, we propose a unifying definition of causality at extreme levels. This definition does not rely on interventionist notions and is unifying in the sense that it encompasses tail causality induced by the restricted model class of structural equations models for extremes, i.e., with heavy-tailed errors, as well as tail causality induced by asymmetric strong tail dependencies. One advantage of our approach is the use of the asymptotically-motivated multivariate Generalized Pareto distribution to properly rescale the random vector of interest. This rescaling, in conjunction with the asymmetric nature of tail causality—whereby extreme events in the cause lead to extreme events in the effect, but not necessarily vice versa—facilitates the identification of the direction of causality when possible.

When applied to simulated hydrological variables in a network of catchments in Switzerland, our methodology unveiled tail causal relationships between precipitation, discharge, and snowmelt in high-elevation catchments. While the ground truth under extreme conditions remains somewhat ambiguous, our findings are aligned with the expected outcomes derived from physical laws. In low-elevation catchments, the interplay between the studied hydrological variables is more intricate which resulted in more uncertainty when estimating the causal score. This emphasizes the need for expert knowledge when it comes to causal notions. For instance, while identifiable outcomes from causal discovery should be examined by domain experts or validated by laws of nature, unidentifiable outcomes might point towards data-related issues. Although not its primary use, our methodology can thus be used to detect limitations in meteorological simulations where the ground truth is dictated by physical laws.

\newpage
\section*{Appendix: Proof of Proposition~\ref{prop.1}}
\label{sec:appendix}
\begin{proof}
Based on the definition of the Wasserstein distance, one can find an optimal coupling $\pi^{*}=(E', X'_i, \pi)$ such that
\begin{eqnarray}
W_1(X_i,E) = {\rm{E}}_{\pi} \vert E' - X'_i \vert, \notag
\end{eqnarray}
where $E' \overset{d}{=} E$ and $X'_i \overset{d}{=} X_i$. We now rely on the representation~\eqref{StandardPareto} and the fact that $E$ stochastically dominates $X_i$, i.e., $F_{X_i}(t) \geq F_E(t), \forall t \in \mathbb{R}$ to show that
$$t= F_{X_i}^{-1} \{ F_{X_i}(t) \} \geq F^{-1}_{X_i} \{ F_E(t) \}$$
and hence that $X'_i = F^{-1}_{X_i} \{ F_E(E') \} \leq E'$. Thus, we have showed that the optimal coupling $\pi^{*}$ satisfies the following property
\begin{equation*}
    X'_i = F^{-1}_{X_i} \{ F_E(E) \} \leq E',
\end{equation*}
and that $W_1(X_i,E) = {\rm{E}}_{\pi} ( E' - X'_i)$. The desired equivalence is then straightforward
\begin{equation*}
     W_1(X_i, E) \leq W_1(X_j,E) \Leftrightarrow W_1(X_i, E) - W_1(X_j,E) = {\rm{E}} (X_j) - {\rm{E}} (X_i) \leq 0.
\end{equation*}

\end{proof}
\newpage
\bibliographystyle{agsm}  
\bibliography{mybib.bib}

@article{Fatichi2015,
  title = {An overview of current applications, challenges, and future trends in distributed process-based models in hydrology},
  author = {Fatichi, Simone and Zeeman, Matthias J. and Fuhrer, Jürg and Burlando, Paolo},
  journal = {Journal of Hydrology},
  volume = {529},
  pages = {849--859},
  year = {2015},
  publisher = {Elsevier},
  doi = {10.1016/j.jhydrol.2015.05.049}
}

@article{Engelke2020,
  title = {Graphical models for extremes (with discussion)},
  author = {Engelke, Sebastian and Hitz, Alexander S.},
  journal = {Journal of the Royal Statistical Society: Series B (Statistical Methodology)},
  volume = {82},
  pages = {871--932},
  year = {2020},
  publisher = {Wiley Online Library},
  doi = {10.1111/rssb.12351}
}

@article{Dettinger2011,
  title={{Climate change, atmospheric rivers, and floods in California--a multimodel analysis of storm frequency and magnitude changes}},
  author={Dettinger, Michael D},
  journal={Journal of the American Water Resources Association},
  volume={47},
  pages={514--523},
  year={2011},
  publisher={Wiley Online Library}
}

@article{naveau_statistical_2020,
	title = {Statistical {methods} for {extreme} {event} {attribution} in {climate} {science}},
	doi = {https://doi.org/10.1146/annurev-statistics-031219-041314},
	urldate = {2022-11-09},
	journal = {Annual Reviews of Statistics and its Application},
    volume = {7},
    pages={89--110},
	author = {Naveau, Philippe and Hannart, Alexis and Ribes, Aurélien},
	year = {2020}
}

@article{Berghuijs,
author = {Berghuijs, Wouter R. and Woods, Ross A. and Hutton, Christopher J. and Sivapalan, M.},
title = {{Dominant flood generating mechanisms across the United States}},
journal = {Geophysical Research Letters},
volume = {43},
pages = {4382-4390},
doi = {https://doi.org/10.1002/2016GL068070},
url = {https://agupubs.onlinelibrary.wiley.com/doi/abs/10.1002/2016GL068070},
eprint = {https://agupubs.onlinelibrary.wiley.com/doi/pdf/10.1002/2016GL068070},
year = {2016}
}

@article{deidda, 
author = {Deidda, C. and Engelke, S. and De Michele, C.},
title = {Asymmetric Dependence in Hydrological Extremes},
journal = {Water Resources Research},
volume = {59},
year = {2023},
url = {https://doi.org/10.1029/2023WR034512}
}

@article{Ioannis2016,
title = {Conditional independence among max-stable laws},
journal = {Statistics \& Probability Letters},
volume = {108},
pages = {9-15},
year = {2016},
issn = {0167-7152},
doi = {https://doi.org/10.1016/j.spl.2015.08.008},
url = {https://www.sciencedirect.com/science/article/pii/S0167715215002874},
author = {Ioannis Papastathopoulos and Kirstin Strokorb}
}

@article{Tawn1990,
 author = {Jonathan A. Tawn},
 journal = {Biometrika},
 pages = {245--253},
 publisher = {[Oxford University Press, Biometrika Trust]},
 title = {Modelling Multivariate Extreme Value Distributions},
 urldate = {2024-04-25},
 volume = {77},
 year = {1990}
}

@techreport{Schirmer23,
  author      = "Schirmer, M. and Jonas, T.",
  title       = "Enhance existing {Swiss} precipitation products with particular regards to snowfall",
  institution = "Meteoswiss",
  year        = "2023",
url= "https://www.meteoswiss.admin.ch/dam/jcr:75a32bf0-c634-476f-a667-7dc3e23a3c8c/Final-report-Enhance-existing-Swiss-precipitation-products-with-particular-regards-to-snowfall.pdf"
}

@incollection{chavezMhalla24,
  author    = "Valérie Chavez-Demoulin and Linda Mhalla",
  title     = "Causality and extremes",
  publisher = "Chapman~\& Hall / CRC",
  booktitle = "Handbook on Statistics of Extremes",
  year      = "2024",
  editor    = "M.~de Carvalho and R.~Huser and P.~Naveau and B.J.~Reich",
  chapter   = "19",
}

@incollection{NaveauSegers24,
  author    = "Philippe Naveau and Johan  Segers",
  title     = "Multivariate extreme value theory",
  publisher = "Chapman~\& Hall / CRC",
  booktitle = "Handbook on Statistics of Extremes",
  year      = "2024",
  editor    = "M.~de Carvalho and R.~Huser and P.~Naveau and B.J.~Reich",
  chapter   = "7",
}

@article{davisonSmith90, 
author = {Anthony C. Davison and Richard L. Smith},
title = {Models for Exceedances over High Thresholds},
journal = {Journal of the Royal Statistical Society: Series B (Statistical Methodology)},
volume = {52},
year = {1990},
pages = {393--442}
}

@article{rootzen18, 
author = {Holger Rootzén and Johan Segers and Jenny Wadsworth},
title = {Multivariate peaks over thresholds models},
journal = {Extremes},
volume = {21},
year = {2018},
pages = {115--145}
}

@article{rootzen06, 
author = {Holger Rootzén and Nader Tajvidi},
title = {{Multivariate generalized Pareto distributions}},
journal = {Bernoulli},
volume = {12},
year = {2006},
pages = {917--930}
}

@article{Gnecco_2021,
	author = {Nicola Gnecco and Nicolai Meinshausen and Jonas Peters and Sebastian Engelke},
	journal = {The Annals of Statistics},
	pages = {1755 -- 1778},
	title = {{Causal discovery in heavy-tailed models}},
	volume = {49},
	year = {2021}
}

@article{Brunner2019,
author = {Brunner, Manuela I. and Hingray, Benoît and Zappa, Massimiliano and Favre, Anne-Catherine},
title = {Future Trends in the Interdependence Between Flood Peaks and Volumes: Hydro-Climatological Drivers and Uncertainty},
journal = {Water Resources Research},
volume = {55},
pages = {4745-4759},
doi = {https://doi.org/10.1029/2019WR024701},
url = {https://agupubs.onlinelibrary.wiley.com/doi/abs/10.1029/2019WR024701},
year = {2019}
}

@article{Gissibl2018,
author = {Nadine Gissibl and Claudia Kl{\"u}ppelberg},
title = {{Max-linear models on directed acyclic graphs}},
volume = {24},
journal = {Bernoulli},
pages = {2693 -- 2720},
year = {2018},
doi = {10.3150/17-BEJ941},
url = {https://doi.org/10.3150/17-BEJ941}
}

@article{Viviroli2009,
title = {An introduction to the hydrological modelling system {PREVAH} and its pre- and post-processing-tools},
journal = {Environmental Modelling \& Software},
volume = {24},
pages = {1209-1222},
year = {2009},
doi = {https://doi.org/10.1016/j.envsoft.2009.04.001},
author = {D. Viviroli and M. Zappa and J. Gurtz and R. Weingartner}
}

@article{Froidevaux2015,
author = {Froidevaux, P. and Schwanbeck, J. and Weingartner, R. and Chevalier, C. and Martius, O.},
title = {Flood triggering in {Switzerland}: the role of daily  to monthly preceding precipitation},
journal = {Hydrology and Earth System Sciences},
volume = {19},
year = {2015},
pages = {3903--3924},
doi = {10.5194/hess-19-3903-2015}
}

@article{Ferreira2014,
author = {Ana Ferreira and Laurens de Haan},
title = {{The generalized Pareto process; with a view towards application and simulation}},
volume = {20},
journal = {Bernoulli},
pages = {1717 -- 1737},
year = {2014},
doi = {10.3150/13-BEJ538}
}

@misc{R,
address = {Vienna, Austria},
author = {{R Core Team}},
institution = {R Foundation for Statistical Computing},
title = {{R: A Language and Environment for Statistical Computing}},
url = {https://www.r-project.org/},
year = {2017}
}

@article{Kiriliouk2018,
author = {Anna Kiriliouk and Holger Rootzen and Johan Segers and Jennifer L. Wadsworth},
title = {Peaks Over Thresholds Modeling With Multivariate Generalized {Pareto} Distributions},
journal = {Technometrics},
volume = {61},
pages = {123--135},
year  = {2018},
publisher = {Taylor & Francis},
doi = {10.1080/00401706.2018.1462738},
URL = {https://doi.org/10.1080/00401706.2018.1462738}
}

@book{monge1781,
  author = {Monge, Gaspard},
  publisher = {De l'Imprimerie Royale},
  title = {M{\'e}moire sur la th{\'e}orie des d{\'e}blais et des remblais},
  year = 1781
}

@book{villani2008,
  title={Optimal Transport: Old and New},
  author={Villani, C.},
  isbn={9783540710509},
  lccn={2008932183},
  series={Grundlehren der mathematischen Wissenschaften},
  url={https://books.google.ch/books?id=hV8o5R7\_5tkC},
  year={2008},
  publisher={Springer Berlin Heidelberg}
}
\end{document}